\documentclass[fleqn,twoside]{article}
\usepackage{espcrc2}

\usepackage{graphicx}
\usepackage[figuresright]{rotating}

\newcommand{\AmS}{{\protect\the\textfont2
  A\kern-.1667em\lower.5ex\hbox{M}\kern-.125emS}}

\title{The TESLA Time Projection Chamber}

\author{N. Ghodbane\address[DESY]{Deutsches Elektronen Synchrotron, \\
        Notkestra\ss e 85, 22607 Hamburg, Deutschland}\thanks{Speaker
        at the 8$^{th}$ Topical Seminar on Innovative Particle and
        Radiation Detectors Siena, 21-24 October 2002} \\ \textit{(for
        the ECFA-DESY TPC study for a Linear Electron-Positron
        Collider)} }
       
\begin{document}

\begin{abstract}
\vspace{1pc} A large Time Projection Chamber is proposed as part of
the tracking system for a detector at the TESLA electron positron
linear collider.  Different ongoing R\&D studies are reviewed,
stressing progress made on a new type readout technique based on
Micro-Pattern Gas Detectors.
\end{abstract}
\maketitle

\section{Introduction}
TESLA (TeV-Energy Superconducting Linear Accelerator) is an $e^+ e^-$
linear Collider designed to operate in the energy range beyond the
mass of the Z$^0$ and up to the TeV region. If approved, TESLA would
allow a wide physics program, focusing among other things on precision
tests of the established Standard Model, the Higgs mechanisms and the
discovery of new particles.

\noindent
To optimally exploit the physics potential of TESLA, the detector has
to satisfy some stringent requirements. It has to have a very good
track momentum resolution to measure the Z$^0$ recoil mass, an
excellent vertex resolution for heavy flavour identification, a high
resolution in the reconstruction of electromagnetic and hadronic
showers in the calorimeter, and has to be as hermetic as possible. The
general concept of a detector fulfilling these criteria, discussed
within the ECFA-DESY study, is described in details in
\cite{mnich,tesla:tdr}.\\ In this contribution, we describe the
current design of the Time Projection Chamber (TPC) and the different
ongoing studies concerning the readout technologies.

\section{The TESLA central Tracker}\label{sect:tpc}
A large TPC is proposed as the main inner tracker at TESLA.  The
actual design, taking into account the different performance goals, is
described in details in the TESLA TDR~\cite{tesla:tdr}.

\begin{figure}[h!t]
\includegraphics[width=7.5cm]{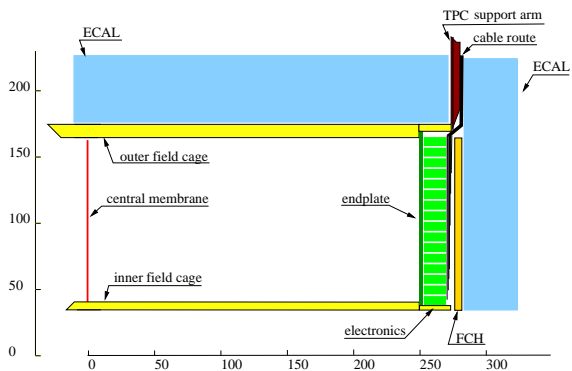}
\caption{General layout of one quarter of the central tracking.}
\label{tpc-side-view.eps}
\end{figure}

\noindent 
The TPC, as shown schematically in figure~\ref{tpc-side-view.eps}, is
of cylindrical design with an inner radius of about 32~cm, an outer
radius of about 170~cm and an overall length in the beam direction of
about 2~$\times$~273~cm. The internal radius is determined by the size
of the mask system, whereas the external radius is constrained by the
fact that the calorimeter is inside the coil. These large dimensions
allow the tracks to be measured with about 200 $(r,\phi,z)$ space
points, so that the global tracking efficiency remains close to 100~\%
in the overall volume and a momentum resolution $\delta\left(
1/p_t\right) < 1.5 \times 10^{-4}$~/~GeV in the central region can be
achieved.  In addition the TPC should measure the specific energy loss
of particles thus aiding in the identification of charged
particles. With Ar-based gas mixtures, operating at atmospheric
pressure, it is expected that a resolution of
$\delta(\log\left(\mbox{dE/dx}\right)) \leq 4.5 \% $ can be reached,
enough, to provide a separation between pions and kaons of better than
two sigma for momenta between 2 and 20 GeV.

\noindent
Operating a TPC in the environment at TESLA presents a number of
challenges. Presently no dedicated hardware trigger is forseen within
a bunch train, so that all detectors should be active throughout one
full train about 1~ms. One readout cycle of the TPC will integrate
over approximatively 150 bunch crossings at TESLA, and many readout
cycles will follow continuously until a train has passed. This makes
it important that the readout is granular enough to resolve the
expected number of hits. In addition means have to be found to avoid
the need for gating the TPC to remove the buildup of positive ions:
during one bunch trains ions will travel significantly into the TPC
drift volume, potentially distorting the drift field.

\noindent 
A possible solution to handle both the high granularity and to suppress the ions drift into the drift volume is the use of Micro-Pattern Gas Detectors as readout devices. Possible candidates are at the moment MicroMEGAS chambers~\cite{MicroMEGAS} or Gas Electron Multipliers (GEM) foils~\cite{GEM}.

\noindent
Micro-Pattern Gas Chambers, due to their intrinsically smaller
distances between amplification elements compared to the traditional
wire chambers, offer the promise to significantly reduce systematic
distortions in the electron drift close to the amplification region
due to the so-called $\vec{\mbox{E}} \times \vec{\mbox{B}}$ effect.

\section{Ongoing Studies}\label{sect:ongoingstudies}
A challenging Research and Development program is underway to meet the
current design goals for a TPC at a linear collider like
TESLA. Institutes from a number of different countries are
participating (Aachen, Carleton, DESY, IPN Orsay, Karlsruhe, Krakow,
LAL, LBNL, MIT, MPI Munich, NIKHEF, Novosibirsk, Rostock, Saclay and
Victoria). The main focus of the work is at the moment on the
development of a valid amplification and ion-suppression scheme (see
reference~\cite{tpc:lcnote}). Over the next years, efforts to develop a
compact readout scheme and to understand the design of the mechanical
structure of such a large TPC are planned. In the following sections
we present the different first results for these different key issues.

\subsection{Gas Amplification Systems}
\begin{figure}[h!t]

\includegraphics[width=3.8cm]{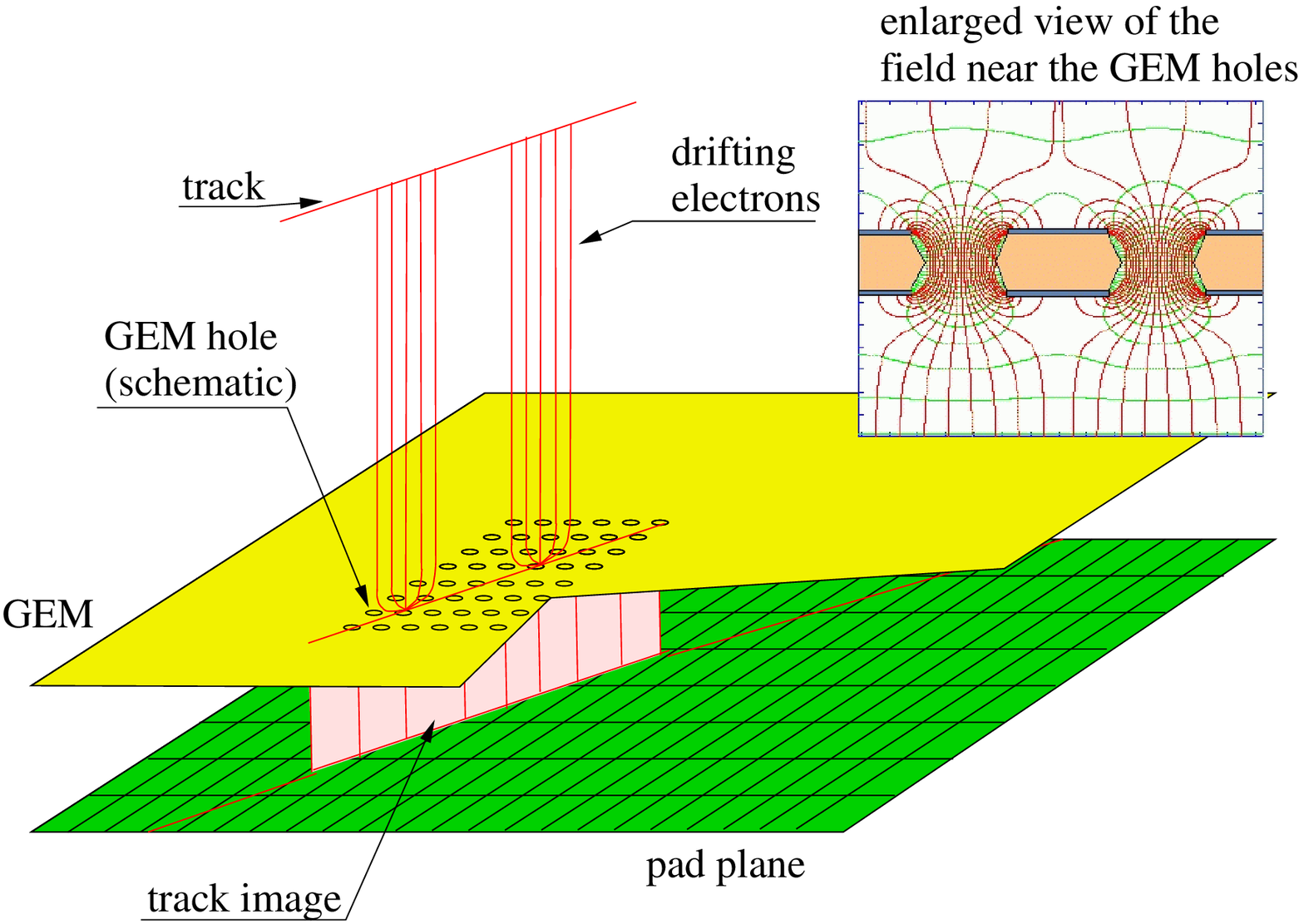}
\includegraphics[width=3.2cm]{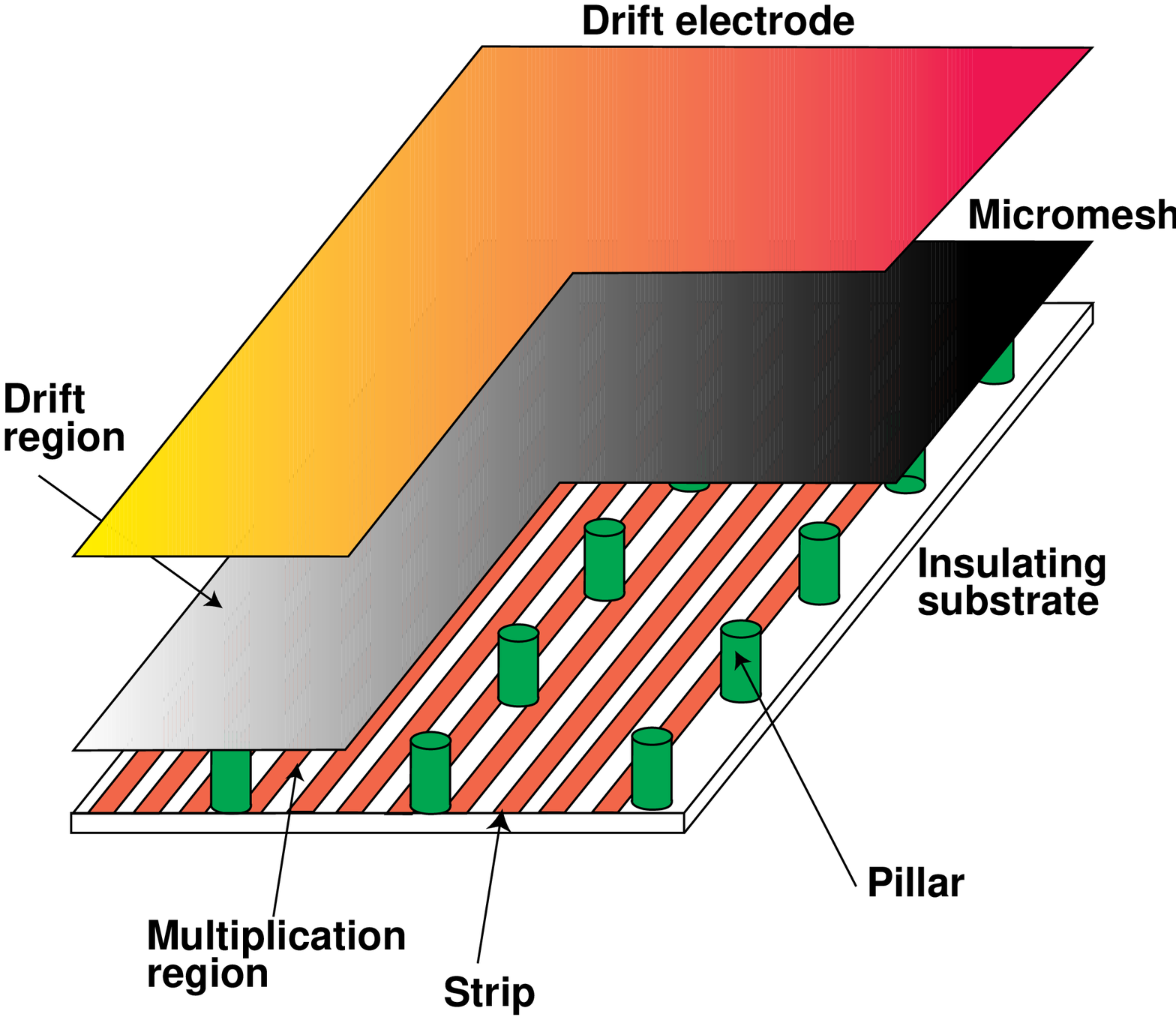}
\caption{Sketch of the GEM (left) and MicroMEGAS.}
\label{gemmicromegas.eps}
\end{figure}

Several studies focus on the TPC readout using Micro-Pattern Gas
Detectors instead of the wire used conventionally to produce the gas
amplification that is necessary to read the TPC signal.  Gas Electron
Multiplier (GEM) and MicroMEGAS, sketched in figure
\ref{gemmicromegas.eps}, are two attractive candidates for the readout
planes.

\noindent
A GEM foil is a sandwich consisting of thin conductive layers of
copper separated by a thin polymer film of capton. The foil is
perforated with small holes arranged in a hexagonal lattice. When
operating in the detector, a potential applied to the two conductive
layers generates a strong electric field, typically 80 kV/cm, in the
small holes.  The electric field lines guide charged particles through
the GEM holes, with minimal losses due to collisions with the
foil. Electrons are accelerated in the strong field lines within the
GEM foil, producing secondary ionizations and charge multiplication
which are then transfered by the electric field to the readout pads
where they are collected.  A unique property of the GEMs is their
capability to operate in cascade, i.e. in a multi-GEM structure,
offering several advantages like a higher gain and an effective
ion feedback suppression.  Most of the groups working with GEM readout
use GEM towers with two or more GEMs.

\noindent
An attractive alternative technology to GEM is that of MicroMEGAS.  An
uniform high-field, typically 30~kV/cm, is produced between a thin
metallic mesh, few microns thick, stretched at a distance of
50-100~$\mu$m above the readout pad plane and held by insulating
pillars.  On top of the mesh, the electrons are transferred to the
amplification gap, where they are multiplied in an avalanche process,
resulting in a large number of electron-ion pairs.  A MicroMEGAS has
similar advantages to a GEM as far as simplicity of construction, low
cost and efficiency of ion-feedback suppression are concerned. It has
an excellent potential for dE/dx resolution due to the fact that the
gain is nearly independent of the gap thickness. Moreover it
is robust and can be built at least in parts using commercially
available components.
\begin{figure}[hb!]
\includegraphics[width=6cm]{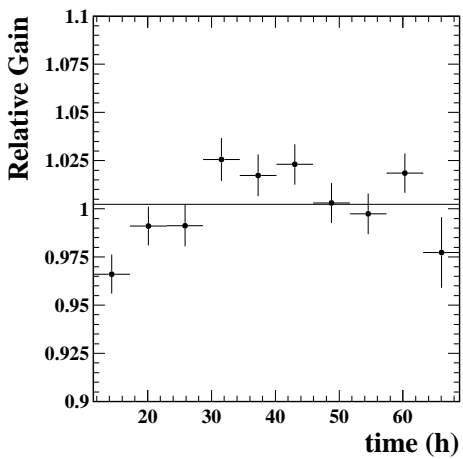}
\caption{Gain uniformity without atmospheric corrections of a two-GEM
structure using the TESLA TDR Gas.}
\label{gainvsruntime.eps}
\end{figure}

\noindent
R\&D studies focus among other things on the gain stability and
homogeneity to reach a dE/dx measurement with 5\% precision. This is
illustrated in figure \ref{gainvsruntime.eps} which shows the gain
uniformity with time using a two-GEM tower structure.

\subsection{Ion Feedback Suppression}
\begin{figure}[h!t]
\includegraphics[width=7.5cm]{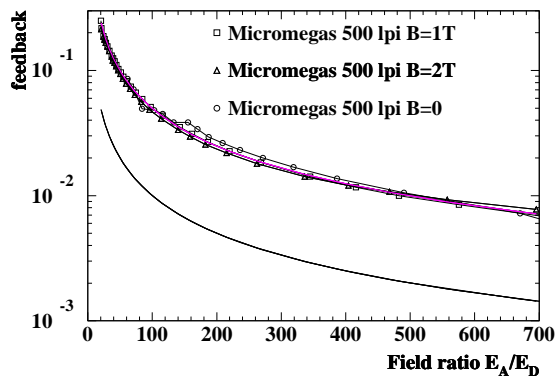}
\caption{Ion feedback suppression using MicroMEGAS for different
electric and magnetic field configurations.}
\label{saclay.eps}
\end{figure}

A crucial issue at TESLA is to prevent the ions produced during the
amplification to return to the TPC drift volume and cause field
distortions. The use of GEMs or MicroMEGAS promises to reduce the ion
feedback. The degree of reduction depends strongly on the exact choice
of operational parameters and the Micro-Pattern Gas Detector
used. Several ongoing studies, using GEMs and MicroMEGAS, demonstrated
a suppression to the level of 10$^{-2}$
(figure~\ref{saclay.eps}). However, a level of suppression to a level
of 10$^{-4}$ is needed to ensure that the positive-ion effects are at
an acceptable level. In case this goal cannot be reached and to
guarantee a stable and robust chamber operation, the insertion of a
gating plane, gated in between bunch trains, might be needed in order
to minimize the effects of ion feedback.

\subsection{Simulations Studies}
To optimally exploit GEMs and MicroMEGAS and in addition to the
experimental program, extensive simulation studies are being
performed. These concern mainly the study of the charge transfer
mechanism in GEM or MicroMEGAS structures. These numerical simulations
are of crucial importance, since they determine the optimal parameter
choice needed to reach the different TESLA TPC milestones.

\begin{figure}[h!]
\includegraphics[width=7.0cm]{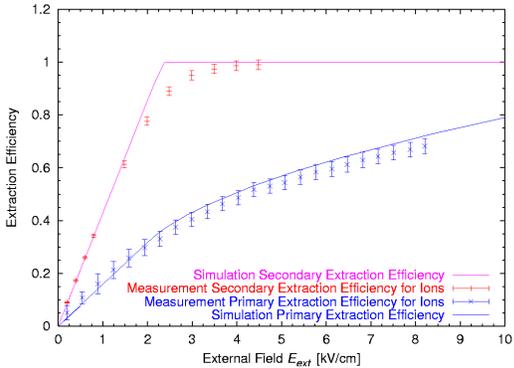}
\caption{Comparison of data and simulation for the ion extraction efficiency from a GEM hole.}\label{aachen.eps}
\end{figure}

\noindent
Figure~\ref{aachen.eps} shows, as the validation of the simulation of
a GEM hole, the comparison between measurement and simulation of the
evolution of the primary and secondary extraction efficiencies
respectively defined as the fraction of ions extracted from the GEM
holes into the drift volume per number of ions produced in the holes
and the fraction of ions extracted from the GEM hole into the drift
volume per number of ions which have been collected in the GEM hole.

\subsection{Resolution and Pad Geometry Studies}
The charge produced during the avalanche in the GEMs is collected by
readout pads located typically a few mm behind the last GEM. Since the
number of pads is limited, the distance between pad centers is large
compared to the size of the electron avalanche cloud (typically a few
to several 100~$\mu$m). The charge from a track after amplification in
the GEMs is therefore sometimes collected on a single pad. This effect
is even more important with magnetic field at TESLA since the
transverse diffusion is then smaller. The major consequence will be
that the centre of gravity method cannot be applied anymore for
position calculation. Ongoing simulation studies investigate more
sophisticated pad geometries, like chevron-shaped pads which may lead
to a better point resolution if the charge sharing between neighboring
pads is more important.
\begin{figure}[h!t]
\includegraphics[width=7.0cm]{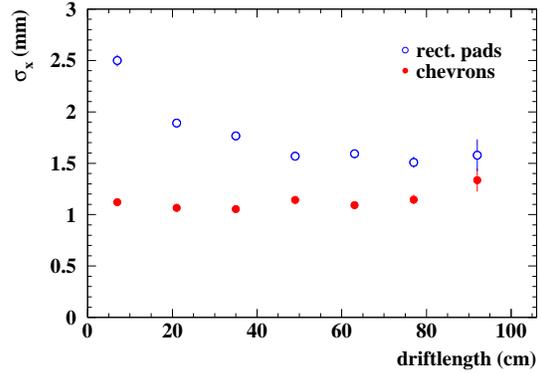}
\caption{Point resolution as a function of the drift length for two
pad geometries, chevrons and rectangular pads.}
\label{res_chevron_rect.2.eps}
\end{figure}

\noindent
Figure~\ref{res_chevron_rect.2.eps} shows the measured point
resolution $\sigma_x$ for rectangular and chevron shaped pad with a
size of 14 $\times$ 14 mm$^2$ and for different drift lengths using
GEMs and without magnetic field. For small drift distances the point
resolution with chevrons is better than with rectangles. This effect
of the pad geometry gets smaller with increasing drift distances since
the influence of the diffusion of the electron cloud gets larger. The
improvement is clearly visible but simulation studies concerning
mainly the pad response function description, the optimal induction
gap between the last GEM and the pad plane, are still required for a
better description of the pad charge sharing mechanism.

\begin{figure}[h!]
\includegraphics[width=7.0cm]{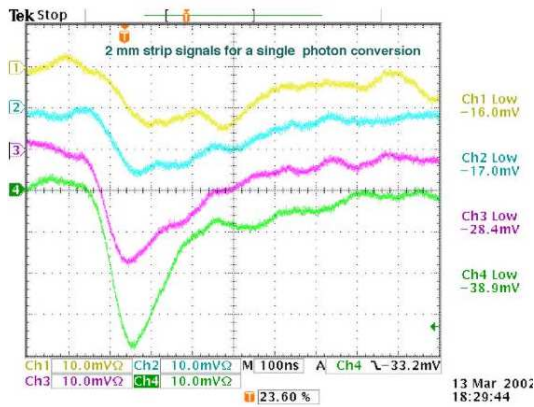}
\caption{A TPC event using a resistive anode GEM.}
\label{carleton.eps}
\end{figure}

\noindent
Another approach to make the electron charge cloud broader, and get
then a better point resolution, uses a resistive anode foil placed a
few ten of $\mu$m between the readout plane and the last GEM. This
work is illustrated by the plot on figure~\ref{carleton.eps} showing
the signal detected with a size of about 2~mm for an initial charge
cluster size of 1~mm.

\subsection{Induced Pulses in a GEM}
The possibility to detect induced pulses in a GEM setup is also being
studied. During the buildup of charges in the GEM hole a small induced
pulse is expected in the pad opposite to this hole. This signal is
also visible in the neighbor pads and can then be used to reconstruct
the position of the avalanche and have a resolution significantly
better. The price to pay for this improvement is that the readout
electronics has to be able to detect the small pulses and that a very
fast electronics is needed. 
\section{Conclusion and Outlook}
Several institutes have started a joint R\&D effort with the goal of
developing the technologies needed for a large TPC at the next
generation of linear colliders. The main challenge is to design a
continuous-tracking, high performance TPC which should be
substantially more performant than any existing TPC. Important
progress has been made to better understand the different aspects of
the TPC. Particular emphasis is placed on the amplification, the
ion-suppression scheme and the large scale-integration electronics.
While first results concerning these different points look very
promising, some open questions like the readout behavior in a strong
magnetic field, or the results with large TPC prototypes, still remain
unanswered and need more investigations in the near future.

\section{Acknowledgments}
I wish to thank all the members of the ECFA-DESY TPC study for
providing material for this contribution. I'm grateful to Ties Behnke
for reading this manuscript.

\end{document}